\documentclass[12pt]{article}
\usepackage{amsmath, amssymb, bm}
\usepackage{booktabs}
\usepackage{multirow}
\usepackage{rotating}
\usepackage{graphicx}
\usepackage[numbers]{natbib}
\usepackage{geometry}
\usepackage{setspace}
\usepackage{authblk}
\usepackage{hyperref}
\usepackage[utf8]{inputenc}
\usepackage{listings}
\usepackage{xcolor}
\usepackage{placeins}

\title{A simulation study to resolve conflicting evidence on the error rates from MANOVA group tests}
\author[1]{Joseph D Consiglio}
\affil[1]{Department of Biotatistics, State University of New York at Buffalo, Buffalo, New York 14214, USA}
\date{}

\begin{document}

\maketitle

\begin{abstract}
Popular software packages report four generalizations of the ANOVA F test when conducting a multivariate analysis of variance (MANOVA). The reported operating characteristics of these fours tests vary widely depending on which research article the reader chooses. Some studies report extremely high type I error rates for a particular test even under ideal assumptions of multivariate normality and homoskedasticity; other studies report rates near the nominal level despite violations of the model assumptions. This simulation study seeks to clarify this apparent contradiction by providing a systematic evaluation of the type I error rates of the four statistics used to test for a group effect in MANOVA.
\end{abstract}


\section{Introduction}
Multivariate analysis of variance (MANOVA) is a tool to accommodate the measurement of more than one response variable on multiple groups of subjects. To test for a significant group effect, the investigator asks whether the observed data are consistant with the assumption of identical group-specific mean response vectors. To carry out this so-called group test in MANOVA, statistical software typically offers four test statistics: Wilks' Lambda, Pillai's Trace, the Hotelling-Lawley Trace, and Roy's Greatest Root. These statistics are discussed in the SAS help file for the ``MANOVA'' statement of \emph{proc glm}, in the R help file for the ``summary.manova'' function, and in section 2. Traditional MANOVA assumes that a hypothetical subject's response vector has a variance-covariance matrix that does not depend on which group the subject belongs to. This is the multivariate analogue of the constant variance assumption in one-way ANOVA.

Ates \emph{et al.} \cite{atecs2019comparison} used simulation to investigate the performance of these four tests under several sample size configurations and several types of violation of normality. They reported that the empirical type I error rate varied somewhat across scenarios, but did not stray far from the nominal level. They stated that, ``Much like ANOVA, the tests are robust to violation of normality and homogeneity of variance when there is balance among the groups.'' Both Sahin \cite{csahin2018monte} and Koc \emph{et al.} \cite{kocc2019monte} simulated data to investigate the type I error properties of the MANOVA test statistics when the responses were not normal, but rather Bernoulli or uniform, obtaining rates that fluctuated within a very small range of the nominal level. Sandamali \cite{sandamali2025eval} used simulation to investigate the type I error properties of the four MANOVA tests in higher dimensions, considering 10 and 20 response variables. Adeleke \emph{et al.} \cite{adeleke2014comparison} investigated the performance of variants of the four MANOVA statistics under non-normality, obtained by truncating their Taylor series expansions; they reported that Roy's Greatest Root was highly robust to violation of normality. With findings contrary to these publications, Adebayo \emph{et al.} \cite{adebayo2018power} examined the four MANOVA tests and reported type I error rates for Roy's Greatest Root reaching 90\% in some scenarios when the model assumptions held. This result is non-intuitive and concerning to the analyst, as one would expect the empirical type I error rate to be somewhat close to the nominal level. Since there are disagreements among the conclusions of the aforementioned authors, this manuscript seeks to carry out a transparent investigation of the type one error properties of the four MANOVA tests.

\section{Materials and Methods}

\subsection{Tests and Notation}
In MANOVA, $r$ $(\geq 2)$ continuous response variables are observed in a $g$-group setting. The $j^{\text{th}}$ value of the $h^{\text{th}}$ response variable measured in group $i$ is $Y_{hij}$, where $h=1,2,\ldots,r$, $i=1,2,\ldots,g$ and $j=1,2,\ldots, n_{i}$. Collecting the $r$ response variables for a given combination of $i$ and $j$ produces the response vector
\[ \underset{(r \times 1)}{\bm{Y}_{ij}} = [Y_{1ij}, Y_{2ij}, \ldots , Y_{rij}]' \]
which has mean vector 
\[ \underset{(r \times 1)}{\bm{\mu}_i}  = [\mu_1, \mu_2, \ldots, \mu_r]'. \]
\noindent The test of interest in MANOVA is whether the $r$-dimensional mean response vectors $\bm{\mu}_i$ are the same for all groups, i.e.
\begin{align*} H_0 \colon	&	\bm{\mu}_1 = \bm{\mu}_2 = \cdots = \bm{\mu}_g \\[ -2ex ]
			  H_a\colon	& \text{not all equal}
\end{align*}

The homogeneity assumption in MANOVA states that the variance-covariance matrix corresponding to the random vector $\bm{Y}_{ij}$ does not depend on group, i.e. $Var(\bm{Y}_{ij}) = \underset{(r \times r)}{\bm{\Sigma}}$ $\forall i$

Define $\bm{Y}_i \in \mathbb{R}^{n_i \times r}$, the matrix of responses in group $i$, as
\begin{equation} 
	\underset{(n_i \times r)}{\bm{Y}_i} = \left[ \begin{array}{cccc}
	Y_{1i1}	& Y_{2i1}	& \cdots	& Y_{ri1} \\
	Y_{1i2}	& Y_{2i2}	& \cdots 	& Y_{ri2} \\
	\vdots 	& \vdots 	& \vdots 	& \vdots \\
	Y_{1in_i}	& Y_{2in_i}	& \cdots 	& Y_{rin_i}
	\end{array} \right].
\label{eq:data_format}
\end{equation}
The mean response vector for group $i$ is
\[ \underset{(r \times 1)}{\bar{\bm{Y}}_i}  = [\bar{Y}_{1i}, \bar{Y}_{2i}, \ldots, \bar{Y}_{ri}]' \]
where
\[ \bar{Y}_{hi} = \frac{1}{n_i} \sum_{j=1}^{n_i} Y_{hij}. \]
The grand mean vector is
\[ \underset{(r \times 1)}{\bar{\bm{Y}}}  = [\bar{Y}_{1}, \bar{Y}_{2}, \ldots, \bar{Y}_{r}]' \]
where
\[ \bar{Y}_{h} = \frac{1}{n_t} \sum_{i=1}^g \sum_{j=1}^{n_i} Y_{hij} \hspace{.2in} \text{and} \hspace{.2in} n_t = \sum_{i=1}^g n_i. \]
The four test statistics used to address the MANOVA hypotheses are built from matrices
\[ \bm{H} = \sum_{i=1}^g n_i (\bar{\bm{Y}}_{i} - \bar{\bm{Y}})(\bar{\bm{Y}}_{i} - \bar{\bm{Y}})' \]
and
\[ \bm{E} = \sum_{i=1}^g \sum_{j=1}^{n_i} ({\bm{Y}}_{ij} - \bar{\bm{Y}}_{i})({\bm{Y}}_{ij} - \bar{\bm{Y}}_{i})' \]
where $\bm{Y}_{ij}$ is a single $r\times 1$ response vector. Here, $\bm{H}$ has been described as the hypothesis sum of squares and cross products matrix, and $\bm{E}$ the error matrix.
Define $\lambda_1, \lambda_2, \ldots, \lambda_r$ to be the eigenvalues of the product $\bm{E}^{-1}\bm{H}$. Then we have
\begin{align*} T_W	& = \prod_{h=1}^r \frac{1}{1+\lambda_h}, 	& \text{\cite{wilks1932certain}} \\
			T_P 		& = \sum_{h=1}^r \frac{\lambda_h}{1+\lambda_h},	& \text{\cite{pillai1955some}} \\
			T_H 	& = \sum_{h=1}^r \lambda_h, 	& \text{\cite{hotelling1931generalization}, \cite{lawley1938generalization}} \\
			T_R 		& = \underset{h}{max} \lambda_h 	& \text{\cite{roy1953heuristic}}
\end{align*}
to denote Wilks' Lambda, Pillai's Trace, the Hotelling-Lawley Trace, and Roy's Greatest Root, respectively.
When computing p-values, SAS and R utilize $F$ approximations for each statistic's null distribution. The transformation
\[ F_W = \frac{1-T_W^{1/s}}{T_W^{1/s}} \cdot \frac{d_2}{d_1} \]
under $H_0$ behaves approximately as an $F$ random variable with $d_1$ and $d_2$ numerator and denominator degrees of freedom, respectively, where
\[ d_1=r(g-1) \hspace{.3in} d_2=n_t - g - r +1 \hspace{.3in} \text{and} \hspace{.3in} s=\sqrt{\frac{r^2(g-1)^2-4}{r^2+(g-1)^2-5}}. \]
The transformations
\[ F_P = \frac{T_P}{(1-T_P)} \cdot \frac{d_2}{d_1} \]
and
\[ F_H = T_H \cdot \frac{d_2}{d_1} \]
also behave as $F_{d_1, d_2}$ random variables under $H_0$.
Roy's Greatest Root differs a bit from the previous three statistics, in that the transformation
\[ F_R = T_R \cdot \frac{d_2}{g} \]
has an approximate $F_{g, d_2}$ null distribution.

\subsection{SAS and R functions}

When applying MANOVA in software, a common option is to organize the data as in \eqref{eq:data_format}, where each column represents a different response variable. We require one additional column to declare the group in which a set of responses occurred. Consider the data structure for an example setting in which we have $r=3$ response variables and $g=3$ groups.
\begin{equation}
	\begin{array}{ccc|c}
	\text{Y1} 	& \text{Y2} 	& \text{Y3} 	& \text{Group} \\
	\hline
	y_{111}	& y_{211}	& y_{311}	& 1 \\
	y_{112}	& y_{212}	& y_{312} 	& 1 \\
	\vdots 	& \vdots 	& \vdots 	& \vdots \\
	y_{11n_1}	& y_{21n_1}	& y_{31n_1} 	& 1 \\
	\vdots 	& \vdots 	& \vdots 	& \vdots \\
	y_{131}	& y_{231}	& y_{331}	& 3 \\
	y_{132}	& y_{232}	& y_{332} 	& 3 \\
	\vdots 	& \vdots 	& \vdots 	& \vdots \\
	y_{13n_3}	& y_{23n_3}	& y_{33n_3} 	& 3
	\end{array}
\label{eq:example_data}
\end{equation}

Applying MANOVA in SAS (version 9.4; SAS Institute Inc., Cary, NC, USA) using a data set structured as in \eqref{eq:example_data} can be done with:
\begin{lstlisting}
proc glm data=data;
	class group;
	model Y1 Y2 Y3 = group / nouni;
	manova h=group;
run;
\end{lstlisting}

The following R (version 4.4.3) code carries out an identical procedure:
\begin{lstlisting}
manova(cbind(data$Y1, data$Y2, data$Y3) ~ factor(data$group))
\end{lstlisting}

In both software packages, the variable ``group'' is declared as a categorical variable, using the \emph{class} statement in SAS and using the \emph{factor} wrapper in R. The consequences of failing to treat the grouping variable as categorical are addressed in section 4.

\subsection{Simulations}

Some of the published results mentioned in the introduction claim that the four MANOVA statistics achieve type I error rates very close to the nominal level under certain scenarios. The first set of simulations conducted here will mirror the scenarios considered in one such paper (\cite{atecs2019comparison}), and show that Roy's Greatest Root does not produce a test whose type I error rate behaves as the other three multivariate tests, but rather is a fair bit larger.

The introductory section mentions another published result (\cite{adebayo2018power}) that makes the fantastic claim that the Roy's Greatest Root test produces simulated type I error on the order of 90\% in certain scenarios. Those scenarios will be reproduced here to demonstrate that the type I error rate for Roy's test is indeed higher than the desired 5\% level, but is not nearly as large as 90\%.

Simulations conducted in R for this manuscript involve 10,000 realizations of the random vector $\underset{(r \times 1)}{\bm{Y}_{ij}}$ choosing $r=3$ response variables and $g=3$ groups. The number of subjects simulated in these groups are denoted by sample size configuration $(n_1, n_2, n_3)$ in the upcoming tables. The group-specific mean vectors were chosen as $[0, 0, 0]$ to allow for assessment of the type I error rate upon 10,000 executions of the MANOVA procedure.

\section{Results}

Table~\ref{tab:Ates3_rep_cat} examines the same sample size settings considered in Ates \emph{et al.} \cite{atecs2019comparison}, simulating $\bm{Y}_{ij} \sim N_3(\bm{0}, \bm{\Sigma})$ where $\bm{\Sigma}$ is taken to be either $\bm{I}_3$ (``homogeneous variance'') or the following matrix with unequal diagonal elements (``heterogeneous variance'')
\[ \bm{\Sigma} =  \left[ \begin{array}{ccc}
	1	& 0	& 0 \\
	0	& 4	& 0 \\
	0	& 0	& 9
	\end{array} \right] \]
for all groups. The grouping variable is treated as categorical, following the syntax detailed for the \emph{manova} function in section 2.2.

\begin{table}[h!]
\centering
\begin{tabular}{|c|c|c|c|c|c|}
\hline
\textbf{Number of responses ($r=3$)} & \textbf{Sample sizes ($n_1$-$n_2$-$n_3$)} & \textbf{Roy} & \textbf{Pillai} & \textbf{Lawley} & \textbf{Wilks} \\
\hline
\multicolumn{6}{|l|}{\textbf{Homogeneous variance}} \\
\hline
& (10-10-10) & 0.1652 & 0.0450 & 0.0573 & 0.0515 \\
& (20-20-20) & 0.1645 & 0.0496 & 0.0552 & 0.0525 \\
& (50-50-50) & 0.1609 & 0.0483 & 0.0499 & 0.0494 \\
& (10-10-20) & 0.1669 & 0.0468 & 0.0526 & 0.0504 \\
& (10-10-50) & 0.1585 & 0.0500 & 0.0533 & 0.0515 \\
& (10-20-20) & 0.1592 & 0.0456 & 0.0510 & 0.0481 \\
& (10-20-50) & 0.1564 & 0.0471 & 0.0500 & 0.0483 \\
& (10-50-50) & 0.1627 & 0.0490 & 0.0504 & 0.0501 \\
& (20-20-50) & 0.1618 & 0.0469 & 0.0504 & 0.0490 \\
& (20-50-50) & 0.1564 & 0.0484 & 0.0506 & 0.0497 \\
\hline
\multicolumn{6}{|l|}{\textbf{Heterogeneous variance}} \\
\hline
& (10-10-10) & 0.1658 & 0.0418 & 0.0515 & 0.0487 \\
& (20-20-20) & 0.1630 & 0.0476 & 0.0532 & 0.0506 \\
& (50-50-50) & 0.1632 & 0.0465 & 0.0487 & 0.0479 \\
& (10-10-20) & 0.1567 & 0.0429 & 0.0495 & 0.0462 \\
& (10-10-50) & 0.1588 & 0.0500 & 0.0544 & 0.0529 \\
& (10-20-20) & 0.1628 & 0.0495 & 0.0549 & 0.0523 \\
& (10-20-50) & 0.1643 & 0.0473 & 0.0514 & 0.0492 \\
& (10-50-50) & 0.1607 & 0.0490 & 0.0511 & 0.0502 \\
& (20-20-50) & 0.1578 & 0.0468 & 0.0499 & 0.0479 \\
& (20-50-50) & 0.1562 & 0.0486 & 0.0511 & 0.0501 \\
\hline
\end{tabular}
\caption{Type I error rates for normal data, $r=3$ responses, treating ``group'' as categorical.}
\label{tab:Ates3_rep_cat}
\end{table}

The striking feature of table~\ref{tab:Ates3_rep_cat} is the noticeably higher type I error rate for the test based on Roy's Greatest Root. While the remaining tests produce rejection rates close to 0.05, the rate for Roy is roughly 0.16 for both choices of $\bm{\Sigma}$ and across all sample size configurations considered.

\begin{table}[h!]
\centering
\begin{tabular}{|c|c|c|c|c|c|}
\hline
\textbf{Number of responses ($r=3$)} & \textbf{Sample sizes ($n_1$-$n_2$-$n_3$)} & \textbf{Roy} & \textbf{Pillai} & \textbf{Lawley} & \textbf{Wilks} \\
\hline
\multicolumn{6}{|l|}{\textbf{Homogeneous variance}} \\
\hline
& (10-10-10) & 0.0499 & 0.0499 & 0.0499 & 0.0499 \\
& (20-20-20) & 0.0464 & 0.0464 & 0.0464 & 0.0464 \\
& (50-50-50) & 0.0502 & 0.0502 & 0.0502 & 0.0502 \\
& (10-10-20) & 0.0541 & 0.0541 & 0.0541 & 0.0541 \\
& (10-10-50) & 0.0514 & 0.0514 & 0.0514 & 0.0514 \\
& (10-20-20) & 0.0501 & 0.0501 & 0.0501 & 0.0501 \\
& (10-20-50) & 0.0482 & 0.0482 & 0.0482 & 0.0482 \\
& (10-50-50) & 0.0469 & 0.0469 & 0.0469 & 0.0469 \\
& (20-20-50) & 0.0503 & 0.0503 & 0.0503 & 0.0503 \\
& (20-50-50) & 0.0533 & 0.0533 & 0.0533 & 0.0533 \\
\hline
\multicolumn{6}{|l|}{\textbf{Heterogeneous variance}} \\
\hline
& (10-10-10) & 0.0487 & 0.0487 & 0.0487 & 0.0487 \\
& (20-20-20) & 0.0490 & 0.0490 & 0.0490 & 0.0490 \\
& (50-50-50) & 0.0512 & 0.0512 & 0.0512 & 0.0512 \\
& (10-10-20) & 0.0493 & 0.0493 & 0.0493 & 0.0493 \\
& (10-10-50) & 0.0476 & 0.0476 & 0.0476 & 0.0476 \\
& (10-20-20) & 0.0518 & 0.0518 & 0.0518 & 0.0518 \\
& (10-20-50) & 0.0479 & 0.0479 & 0.0479 & 0.0479 \\
& (10-50-50) & 0.0514 & 0.0514 & 0.0514 & 0.0514 \\
& (20-20-50) & 0.0501 & 0.0501 & 0.0501 & 0.0501 \\
& (20-50-50) & 0.0486 & 0.0486 & 0.0486 & 0.0486 \\
\hline
\end{tabular}
\caption{Type I error rates for normal data, $r=3$ responses, treating ``group'' as continuous.}
\label{tab:Ates3_rep_cont}
\end{table}

Table~\ref{tab:Ates3_rep_cont} uses the same data generated for table~\ref{tab:Ates3_rep_cat}, but uses a continuous interpretation of the grouping variable by removing the \emph{factor} wrapper when conducting the MANOVA analysis. The results are conspicuous: Roy's Greatest Root no longer produces a test with inflated type I error. Indeed, all four multivariate tests produce rates in very close proximity to 0.05 across scenarios. Compare these rates to those presented in table~\ref{tab:Ates3} drawn from \cite{atecs2019comparison}. It becomes apparent that the authors may have inadvertantly treated ``group'' as continuous.

\begin{table}[h!]
\centering
\begin{tabular}{|c|c|c|c|c|c|}
\hline
\textbf{Number of responses: $r=3$} & \textbf{Sample sizes ($n_1$-$n_2$-$n_3$)} & \textbf{Roy} & \textbf{Pillai} & \textbf{Lawley} & \textbf{Wilks} \\
\hline
\multicolumn{6}{|l|}{\textbf{Homogeneous variance}} \\
\hline
& (10-10-10) & 0.0485 & 0.0504 & 0.0484 & 0.0476 \\
& (20-20-20) & 0.0519 & 0.0454 & 0.0489 & 0.0516 \\
& (50-50-50) & 0.0467 & 0.0479 & 0.0497 & 0.0516 \\
& (10-10-20) & 0.0504 & 0.0519 & 0.0495 & 0.0511 \\
& (10-10-50) & 0.0517 & 0.0486 & 0.0522 & 0.0519 \\
& (10-20-20) & 0.0477 & 0.0475 & 0.0518 & 0.0511 \\
& (10-20-50) & 0.0527 & 0.0468 & 0.0459 & 0.0483 \\
& (10-50-50) & 0.0501 & 0.0464 & 0.0529 & 0.0498 \\
& (20-20-50) & 0.0467 & 0.0504 & 0.0504 & 0.0503 \\
& (20-50-50) & 0.0526 & 0.0495 & 0.0527 & 0.0500 \\
\hline
\multicolumn{6}{|l|}{\textbf{Heterogeneous variance}} \\
\hline
& (10-10-10) & 0.0497 & 0.0494 & 0.0497 & 0.0520 \\
& (20-20-20) & 0.0539 & 0.0493 & 0.0521 & 0.0476 \\
& (50-50-50) & 0.0497 & 0.0478 & 0.0468 & 0.0526 \\
& (10-10-20) & 0.0498 & 0.0476 & 0.0511 & 0.0488 \\
& (10-10-50) & 0.0486 & 0.0524 & 0.0528 & 0.0492 \\
& (10-20-20) & 0.0501 & 0.0477 & 0.0495 & 0.0544 \\
& (10-20-50) & 0.0507 & 0.0492 & 0.0467 & 0.0531 \\
& (10-50-50) & 0.0500 & 0.0493 & 0.0528 & 0.0481 \\
& (20-20-50) & 0.0478 & 0.0534 & 0.0551 & 0.0478 \\
& (20-50-50) & 0.0503 & 0.0476 & 0.0541 & 0.0533 \\
\hline
\end{tabular}
\caption{Results from Ates \emph{et al.} - Type I error rates for normal data, $r=3$ responses.}
\label{tab:Ates3}
\end{table}

\FloatBarrier

Within any given row of table~\ref{tab:Ates3_rep_cont}, we see identical type I error rates for the four multivariate tests. To demonstrate why this occurs, consider table ~\ref{tab:Routput}, which displays R output from executing the manova(.) function on a data set having $g=3$ and $(n_1, n_2, n_3)=(20, 20, 10)$ for the categorical treatment of group, followed by the continuous treatment. SAS output follows likewise in table~\ref{tab:SASoutput}. 

\begin{table}[h!]
\centering
\begin{tabular}{l | llllll}
\toprule
	\textbf{Group} 		& \textbf{Statistic}		& \textbf{Value} 	& \textbf{F Value} 	& \textbf{Num DF} 	& \textbf{Den DF}		& \textbf{Pr $>$ F} \\
\midrule
	\multirow{4}{*}{\textbf{Categorical}}
	& \text{Wilks' Lambda}			& 0.9005 		& 0.8070 		& 6 				& 90				& 0.5672 \\
	& \text{Pillai's Trace} 			& 0.1002 		& 0.8091 			& 6 				& 92				& 0.5655 \\
	& \text{Hotelling-Lawley Trace} 	& 0.1097 		& 0.8042 			& 6 				& 88				& 0.5693 \\
	& \text{Roy's Greatest Root}		& 0.1015 		& 1.5566 		& 3 				& 46				& 0.2127 \\
\midrule
	\multirow{4}{*}{\textbf{Continuous}}
	& \text{Wilks' Lambda}		& 0.9847 		& 0.2388 			& 3 				& 46				& 0.8688 \\
	& \text{Pillai's Trace} 			& 0.0153 		& 0.2388 			& 3 				& 46				& 0.8688 \\
	& \text{Hotelling-Lawley Trace} 	& 0.0156 		& 0.2388 			& 3 				& 46				& 0.8688 \\
	& \text{Roy's Greatest Root}	& 0.0156 		& 0.2388 			& 3 				& 46				& 0.8688 \\
\bottomrule
\end{tabular}
\caption{R output, MANOVA applied to a fixed data set}
\label{tab:Routput}
\end{table}

\begin{table}[h!]
\centering
\begin{tabular}{l | llllll}
\toprule
	\textbf{Group} 		& \textbf{Statistic}		& \textbf{Value} 	& \textbf{F Value} 	& \textbf{Num DF} 	& \textbf{Den DF}		& \textbf{Pr $>$ F} \\
\midrule
	\multirow{4}{*}{\textbf{Categorical}}
	& \text{Wilks' Lambda}			& 0.9005 		& 0.81 			& 6 				& 90				& 0.5672 \\
	& \text{Pillai's Trace} 			& 0.1002 		& 0.81 			& 6 				& 92				& 0.5655 \\
	& \text{Hotelling-Lawley Trace} 	& 0.1097 		& 0.81 			& 6 				& 58.256			& 0.5635 \\
	& \text{Roy's Greatest Root}		& 0.1015 		& 1.56 			& 3 				& 46				& 0.2127 \\
\midrule
	\multirow{4}{*}{\textbf{Continuous}}
	& \text{Wilks' Lambda}			& 0.9847 		& 0.24 			& 3 				& 46				& 0.8688 \\
	& \text{Pillai's Trace} 			& 0.0153 		& 0.24 			& 3 				& 46				& 0.8688 \\
	& \text{Hotelling-Lawley Trace} 	& 0.0156 		& 0.24 			& 3 				& 46				& 0.8688 \\
	& \text{Roy's Greatest Root}		& 0.0156 		& 0.24 			& 3 				& 46				& 0.8688 \\
\bottomrule
\end{tabular}
\caption{SAS output, MANOVA applied to a fixed data set}
\label{tab:SASoutput}
\end{table}

\FloatBarrier

The "F Value" representations (detailed in section 2.1) of all four statistics are identical. Continuous treatment of group produces results for which all four tests must jointly reject $H_0$ or fail to reject $H_0$ with no opportunity for disagreement. This behavior causes the empirical type I error rates to coincide exactly for all four statistics. When failing to treat the grouping variable as categorical, it is no longer true that multivariate analysis of variance has occured; details appear in section 4.1.

\begin{table}[h!]
\centering
\begin{tabular}{lccccc}
\toprule
	\textbf{Sample Sizes}  	&   			&			&			& \\
	($n_1$-$n_3$-$n_3$)	& \textbf{Wilks} 	& \textbf{Lawley} 	& \textbf{Pillai} 	& \textbf{Roy} \\
\midrule
	(10-10-10)		& 0.053 & 0.056 & 0.047 & 0.167 \\
	(100-100-100) 	& 0.048 & 0.049 & 0.048 & 0.152 \\
	(1000-1000-1000) 	& 0.050 & 0.050 & 0.050 & 0.153 \\
\midrule
	(10-20-30) 		& 0.052 & 0.054 & 0.050 & 0.164 \\
	(100-200-300) 	& 0.050 & 0.050 & 0.050 & 0.157 \\
	(600-800-1000) 	& 0.047 & 0.047 & 0.047 & 0.159 \\
\bottomrule
\end{tabular}
\caption{Type I error rates for normal data, $r=3$ choosing $\bm{\Sigma} = \bm{I}_3$.}
\label{tab:Adebayo5_my_reproduction}
\end{table}

\FloatBarrier

Table~\ref{tab:Adebayo5_my_reproduction}, addresses the scenarios considered in \cite{adebayo2018power} using $r=3$, $g=3$, and obtaining 10,000 realization of $\bm{Y}_{ij}$ from a multivariate normal distribution with zero mean vector and $\bm{\Sigma} = \bm{I}_3$. Several other choices for the common covariance matrix were considered, but produced negligible effect on the observed error rates. The behavior seen here is in very close agreement with table~\ref{tab:Ates3_rep_cat}; we see that three of the multivariate tests incorrectly reject the hypothesis of equal mean vectors roughly 5\% of the time, while the test based on Roy's Greatest Root has an inflated rejection rate of roughly 16\%. Compare this behavior to the results quoted in table~\ref{tab:Adebayo5}, an excerpt from Adebayo \emph{et al.} \cite{adebayo2018power}. The reported values of type I error are extremely large, cresting 90\% for Roy's Greatest Root in the largest sample size configurations. The quoted results also claim that extreme violations of type I error control occur for the remaining three multivariate tests. This behavior is not corroborated here in table~\ref{tab:Adebayo5_my_reproduction}, where Wilks' Lambda, the Hotelling-Lawley trace, and Pillai's trace all produce very reasonable and expected error rates.
We note that the form of the common $\bm{\Sigma}$ is not specified with any detail in \cite{adebayo2018power}; the authors only say that the covariance matrix is held equal across groups.

\begin{table}[h!]
\centering
\begin{tabular}{lccccc}
\toprule
	\textbf{Sample Sizes}  	&   			&			&			& \\
	($n_1$-$n_3$-$n_3$)	& \textbf{Wilks} 	& \textbf{Lawley} 	& \textbf{Pillai} 	& \textbf{Roy} \\
\midrule
	(10-10-10)		& 0.005 & 0.068 & 0.061 & 0.150 \\
	(100-100-100) 	& 0.026 & 0.117 & 0.113 & 0.401 \\
	(1000-1000-1000) 	& 0.481 & 0.834 & 0.834 & 0.967 \\
\midrule
	(10-20-30) 		& 0.003 & 0.068 & 0.079 & 0.284 \\
	(100-200-300) 	& 0.005 & 0.219 & 0.218 & 0.568 \\
	(600-800-1000) 	& 0.225 & 0.713 & 0.715 & 0.936 \\
\bottomrule
\end{tabular}
\caption{Type I error rates for normal data, homogeneous group variance, $r=3$ responses reported in Adebayo.}
\label{tab:Adebayo5}
\end{table}

\FloatBarrier

Finally, to consider violations of the assumption that the group-specific covariance matrices are equal, the error rates observed here are presented in table~\ref{tab:Adebayo6_my_reproduction} for two choices of $\bm{\Sigma}_i$ described as ``Identity'' and ``Toeplitz.'' The label ``Identity'' refers to choosing 

\[ \bm{\Sigma}_1 =  \left[ \begin{array}{ccc}
	1	& 0	& 0 \\
	0	& 1	& 0 \\
	0	& 0	& 1
	\end{array} \right],  \hspace{.2in} \bm{\Sigma}_2 =  \left[ \begin{array}{ccc}
	4	& 0	& 0 \\
	0	& 4	& 0 \\
	0	& 0	& 4
	\end{array} \right], \hspace{.2in} \bm{\Sigma}_3 =  \left[ \begin{array}{ccc}
	9	& 0	& 0 \\
	0	& 9	& 0 \\
	0	& 0	& 9
	\end{array} \right] \]

\noindent while in the section labeled ``Toeplitz,'' the group-specific covariance matrices have the form

\[ 
\bm{\Sigma}_i = \sigma^2 \left[ \begin{array}{ccc}
	1		& \rho_i	& \rho_i^2 \\
	\rho_i 	& 1		& \rho_i \\
	\rho_i^2	& \rho_i	& 1
\end{array} \right]
\]

\noindent with $\rho_1=0.75$, $\rho_2=0.5$, and $\rho_3=0.25$.

\begin{table}[h!]
\centering
\begin{tabular}{l | cccc | cccc}
\toprule
	\textbf{Sample Sizes}  	& \multicolumn{4}{c}{Identity}	 & \multicolumn{4}{c}{Toeplitz} \\
	($n_1$-$n_3$-$n_3$)	& \textbf{Wilks} 	& \textbf{Lawley} 	& \textbf{Pillai} 	& \textbf{Roy} 	& \textbf{Wilks} 	& \textbf{Lawley} 	& \textbf{Pillai} 	& \textbf{Roy}\\
\midrule
	(10-10-10)			& 0.076 & 0.085 & 0.063 & 0.220		& 0.058 & 0.061 & 0.050 & 0.175 \\
	(100-100-100) 		& 0.070 & 0.071 & 0.068 & 0.198 	& 0.052 & 0.052 & 0.052 & 0.160 \\
	(1000-1000-1000) 	& 0.065 & 0.065 & 0.065 & 0.199 	& 0.053 & 0.053 & 0.053 & 0.160  \\
\midrule
	(10-20-30) 		& 0.011 & 0.013 & 0.010 & 0.070 	& 0.037 & 0.039 & 0.035 & 0.131 \\
	(100-200-300) 	& 0.012 & 0.012 & 0.012 & 0.064 	& 0.037 & 0.037 & 0.037 & 0.127 \\
	(600-800-1000) 	& 0.027 & 0.027 & 0.027 & 0.121 	& 0.043  & 0.043 & 0.043 & 0.140 \\
\bottomrule
\end{tabular}
\caption{Type I error rates for normal data, $r=3$, heterogenous group variance.}
\label{tab:Adebayo6_my_reproduction}
\end{table}

\FloatBarrier

Table~\ref{tab:Adebayo6} quotes results for the corresponding scenarios printed in Adebayo \emph{et al.} \cite{adebayo2018power}. The error rates quoted from \cite{adebayo2018power} are clearly much higher than those obtained here, especially regarding Roy's test.

\begin{table}[h!]
\centering
\begin{tabular}{l | cccc}
\toprule
	\textbf{Sample Sizes}  	& \multicolumn{4}{c}{Unspecified Heterogeneous}	 \\
	($n_1$-$n_3$-$n_3$)	& \textbf{Wilks} 	& \textbf{Lawley} 	& \textbf{Pillai} 	& \textbf{Roy} \\
\midrule
	(10-10-10)			& 0.034 & 0.137 & 0.110 & 0.272	\\
	(100-100-100) 		& 0.021 & 0.101 & 0.094 & 0.403 \\
	(1000-1000-1000) 	& 0.052 & 0.168 & 0.168 & 0.493 \\
\midrule
	(10-20-30) 		& 0.005 & 0.112 & 0.099 & 0.331 \\
	(100-200-300) 	& 0.006 & 0.095 & 0.093 & 0.385 \\
	(600-800-1000) 	& 0.013 & 0.141 & 0.141 & 0.456 \\
\bottomrule
\end{tabular}
\caption{Type I error rates for normal data, heterogeneous group variance, $r=3$ responses reported in Adebayo.}
\label{tab:Adebayo6}
\end{table}

\section{Discussion}

\subsection{Regarding categorical response}

Failing to treat ``group'' as a categorical variable does not result in MANOVA; instead software performs a multivariate linear regression according to the following model:
\begin{align*}
	Y_1	& = \beta_{10} + \beta_{11} \cdot group + \varepsilon_1 \\
	Y_2	& = \beta_{20} + \beta_{21} \cdot group + \varepsilon_2 \\
		& \hspace{.8in} \vdots \\
	Y_r 	& = \beta_{r0} + \beta_{r1} \cdot group + \varepsilon_r
\end{align*}
The error vector $\bm{\varepsilon} = [\varepsilon_1, \varepsilon_2, \ldots, \varepsilon_r]'$ is assumed to be such that $\bm{\varepsilon} \sim N_r(\bm{0}, \bm{\Sigma})$ with $\bm{\Sigma}$ unstructured and having homoskedasticity in a multivariate sense (i.e. any response vector has covariance matrix $\bm{\Sigma}$, regardless of which group the associated subject belongs to).
The hypotheses tested using the four MANOVA statistics no longer address equality of group-specific mean vectors, since ``group'' has now been interpreted as a continuous predictor variable. Instead, the MANOVA tests address whether the $r$ slope coefficients are simultaneously zero, as stated below.
\begin{align*} H_0 \colon	&	\beta_{11} = \beta_{21} = \cdots = \beta_{r1} = 0 \\[ -2ex ]
		H_a\colon	& \text{not all zero}
\end{align*}
When investigating the type I error properties of the four MANOVA tests, we require equality of mean response vectors across groups. It is convenient and natural to simulate realizations of the response vector $\bm{Y}$ from a multivariate distribution with mean $\bm{0}$. If the program to be run in software does not declare the grouping variable to be categorical, the data are simulated within the scope of the null hypothesis for multivariate linear regression. The analyst would expect to reject the null hypothesis that the vector of slope coefficients is equal to the zero vector in roughly $100 \alpha$ percent of cases. The type I error rates observed in Ates (table~\ref{tab:Ates3}) and reproduced independently here (table~\ref{tab:Ates3_rep_cont}) indicate that this behavior holds regardless of which multivariate test statistic is used; we observe rejection rates that appear to fluctate randomly around 0.05 for all sample size configurations considered.

\subsection{Regarding unreasonable type I error}

Table~\ref{tab:Adebayo5_my_reproduction} indicates that when responses are drawn from multivariate normal distributions with equal covariance matrices across groups, three of the four MANOVA tests achieve type I error rates very close to the nominal 0.05. Roy's Greatest Root is the notable exception, producing a rejection of equality of mean vectors roughly 16\% of the time. While this is unacceptably large, it is nowhere near as large as the 96.7\% rejection rate reported in \cite{adebayo2018power} for the largest sample size case. The extreme rejection rates in table~\ref{tab:Adebayo5} suggest potential errors in the original simulation procedure. The high-dimensional cases considered in \cite{sandamali2025eval} do show type I error rates for Roy's Greatest Root on the order of 90\% when the number of groups increases to 9; perhaps Adebayo \emph{et al.} simulated under such a scenario.

Table~\ref{tab:Adebayo6_my_reproduction} investgates the robustness of the MANOVA tests to violation of the assumption that the group-specific covariance matrices are equal. The general pattern in cases of balanced group sample sizes is that the type I error rates for the Toeplitz case are smaller than when using group-specific diagonal covariance matrices. For imbalanced sample size configurations, the reverse is true: all type I error rates are clearly smaller when using diagonal covariance than their Toeplitz counterparts. With the exception of Roy's Greatest Root, all rates for imbalanced scenarios are noticeably below the nominal level, producing conservative tests. Across all scenarios, use of Roy's Greatest Root is associated with an unacceptably high rejection rate. 

The test based on Roy's Greatest Root has been criticized as early as 1976 by Olsen \cite{olson1976choosing}, who says ``The largest-root test \emph{R} must be rejected by almost any standard. It results in far too many false claims of significance when assumptions are violated.'' The reason for this behavior is related to the crude nature of the test statistic, defined in Section 2 as $T_R$, the largest eigenvalue of the matrix $\bm{E}^{-1}\bm{H}$. While the other three test statistics make use of the full set of eigenvalues $\lambda_1, \lambda_2, \cdots, \lambda_r$, Roy's Greatest Root uses only the limited information present in $\underset{h}{max} \{ \lambda_h \}$, providing only a partial summarization of the underlying data. Again from Olsen, ``The largest root . . . is apt to be both wasteful of information and susceptible to substantial dislocation due to a single deviant eigenvalue.''  Olsen highlights that Roy's test has an unacceptably high type I error probability when the assumptions of MANOVA are violated. The simulations presented here (tables ~\ref{tab:Ates3_rep_cat}, ~\ref{tab:Adebayo5_my_reproduction}) make claims that go further: the issue of inflated type I error exists for Roy's test even under ideal modeling conditions, when the responses in each group are generated from identical multivariate normal distributions.

\section{Conclusion}

The shortcomings of Roy's Greatest Root have been documented half a century ago, but there are recent published results claiming a performance of Roy's Greatest Root on par with that of the other three multivariate tests. This appears to be due to a continuous treatment of the grouping variable, obscuring the fact that Roy's test produces an abundance of type I errors. Other recent results acknowledge the inflated type I error rate of Roy's test, but highly exaggerate the severity of inflation. The simulations here confirm longstanding concerns about Roy’s Greatest Root and emphasize the importance of treating grouping variables as categorical when carrying out MANOVA.

\bibliographystyle{plainnat}

\bibliography{manova_references}

@article{adebayo2018power,
  title={Power and Type I Error Rate Comparison of Multivariate Analysis of Variance},
  author={Adebayo, Patrick and Ibrahim, Ahmed},
  journal={Trends in Science \& Technology Journal},
  volume={3},
  number={2},
  pages={628--635},
  year={2018}
}

@article{adeleke2014comparison,
  title={A Comparison of Some Test Statistics for Multivariate Analysis of Variance Model With Non-Normal Responses},
  author={Adeleke, Babatunde Lateef and Yahaya, WB and Usman, Abubakar},
  year={2014},
  publisher={Journal of Natural Sciences Research}
}

@article{csahin2018monte,
  title={A Monte Carlo Simulation Study Robustness of MANOVA Test Statistics in Bernoulli Distribution},
  author={{\c{S}}ahin, Mustafa and Ko{\c{c}}, {\c{S}}eyma},
  journal={S{\"u}leyman Demirel {\"U}niversitesi Fen Bilimleri Enstit{\"u}s{\"u} Dergisi},
  volume={22},
  number={3},
  pages={1125--1131},
  year={2018},
  publisher={S{\"u}leyman Demirel University}
}

@article{atecs2019comparison,
  title={Comparison of Test Statistics of Nonnormal and Unbalanced Samples for Multivariate Analysis of Variance in terms of Type-I Error Rates},
  author={Ate{\c{s}}, Can and Kaymaz, {\"O}zlem and Kale, H Emre and Tekindal, Mustafa Agah},
  journal={Computational and Mathematical Methods in Medicine},
  volume={2019},
  number={1},
  pages={2173638},
  year={2019},
  publisher={Wiley Online Library}
}

@article{hotelling1931generalization,
  title={The Generalization of Student's Ratio},
  author={Hotelling, Harold and others},
  year={1931},
  publisher={Springer}
}

@article{kocc2019monte,
  title={A Monte Carlo Simulation Study Robustness of Manova Test Statistics in Bernoulli and Uniform Distribution},
  author={Ko{\c{c}}, {\c{S}}eyma and {\c{C}}anga, Demet and {\"O}nem, Ay{\c{s}}e Bet{\"u}l and Yavuz, Esra and {\c{S}}ahin, Mustafa},
  journal={Black Sea Journal of Engineering and Science},
  volume={2},
  number={2},
  pages={42--51},
  year={2019},
  publisher={U{\u{g}}ur {\c{S}}EN}
}

@article{lawley1938generalization,
  title={A Generalization of Fisher's Z Test},
  author={Lawley, Derrick N},
  journal={Biometrika},
  volume={30},
  number={1/2},
  pages={180--187},
  year={1938},
  publisher={JSTOR}
}

@article{olson1976choosing,
  title={On Choosing a Test Statistic in Multivariate Analysis of Variance.},
  author={Olson, Chester L},
  journal={Psychological Bulletin},
  volume={83},
  number={4},
  pages={579},
  year={1976},
  publisher={American Psychological Association}
}

@article{pillai1955some,
  title={Some new test criteria in multivariate analysis},
  author={Pillai, KC Sreedharan},
  journal={The Annals of Mathematical Statistics},
  pages={117--121},
  year={1955},
  publisher={JSTOR}
}

@article{roy1953heuristic,
  title={On a Heuristic Method of Test Construction and its Use in Multivariate Analysis},
  author={Roy, Samarendra Nath},
  journal={The Annals of Mathematical Statistics},
  volume={24},
  number={2},
  pages={220--238},
  year={1953},
  publisher={Institute of Mathematical Statistics}
}

@phdthesis{sandamali2025eval,
  title={Evaluation of MANOVA test statistics for increasing number of groups},
  author={Sandamali, Irosha},
  type={PhD dissertation},
  school={Uppsala University, Department of Statistics},
  year={2025},
  url={https://uu.diva-portal.org/smash/get/diva2:1978126/FULLTEXT01.pdf}
}

@article{wilks1932certain,
  title={Certain Generalizations in the Analysis of Variance},
  author={Wilks, Samuel S},
  journal={Biometrika},
  volume={24},
  number={3/4},
  pages={471--494},
  year={1932},
  publisher={JSTOR}
}

\end{document}